\documentclass[namedreferences]{solarphysics}

\usepackage[optionalrh]{spr-sola-addons} 
\usepackage{graphicx}        
\usepackage{color}           

\begin{document}

\begin{article}

\begin{opening}

\title{Simulations of 3D Magnetic Merging: Resistive Scalings for Null Point and QSL Reconnection}

\author{Frederic~\surname{Effenberger}$^{1,2}$ and I.~J.~D.~\surname{Craig}$^{1}$}
\runningauthor{F.~Effenberger, I.J.D.~Craig}
\runningtitle{Simulations of 3D Magnetic Merging}
\institute{$^{1}$ Department of Mathematics, University of Waikato, Private Bag 3105, Hamilton, New Zealand \\ 
 $^{2}$ Department of Physics and KIPAC, Stanford University, Stanford, CA 94305, USA, feffen@stanford.edu}

\begin{abstract}
  Starting from an exact, steady-state, force-free solution of the
  magnetohydrodynamic (MHD) equations, we investigate how resistive
  current layers are induced by perturbing line-tied three-dimensional
  magnetic equilibria. This is achieved by the superposition of a weak
  perturbation field in the domain, in contrast to studies where the
  boundary is driven by slow motions, like those present in
  photospheric active regions. Our aim is to quantify how the current
  structures are altered by the contribution of so called
  quasi-separatrix layers (QSLs) as the null point is shifted outside
  the computational domain. Previous studies based on
  magneto-frictional relaxation have indicated that, despite the
  severe field line gradients of the QSL, the presence of a null is
  vital in maintaining fast reconnection.  Here, we explore this
  notion using highly resolved simulations of the full MHD
  evolution. We show that for the null-point configuration, the
  resistive scaling of the peak current density is close to
  $J\sim\eta^{-1}$, while the scaling is much weaker, \emph{i.e.}\
  $J\sim\eta^{-0.4}$, when only the QSL connectivity gradients provide
  a site for the current accumulation.
\end{abstract}

\end{opening}

\section{Introduction}
\label{sec:introduction}
The development of current singularities in a three-dimensional
magnetohydrodynamic (MHD) plasma evolution is an active topic in
coronal astrophysics with high relevance to the general problem of
magnetic reconnection \citep{Priest-Forbes-2000} and related particle
acceleration mechanisms (\emph{e.g.}
\opencite{Heerikhuisen-etal-2002}; \opencite{Stanier-etal-2012}). In
particular, if reconnection is to be effective in altering the
magnetic topology, very strong localized currents must be present in
the vicinity of the reconnection site. If these conditions on the
current density are not met, the weak coronal resistivity can strongly
inhibit the reconnection rate.

In the past, models of two-dimensional (2D) reconnection have yielded
much insight into the formation of near singular current sheets, as
supported by 2D MHD simulations \citep[\emph{e.g.}][]{Biskamp-1986}
and analytic modeling \citep[\emph{e.g.}][]{Forbes-1982}. Although the
three-dimensional merging problem has been less intensively explored,
it is known that 3D merging may involve not only current sheets, as in
``fan'' reconnection, but also the quasi-cylindrical current tubes of
``spine'' reconnection \citep[\emph{e.g.}][]{Craig-Fabling-1996}. A
further form, ``separator'' reconnection, involves thin current
ribbons that form along field lines linking any two nulls
\citep{Heerikhuisen-Craig-2004}. \citet{Longcope-2005} gives a review
on the topological aspects of three-dimensional reconnection. These
topological forms derive from the 3D eigenstructure of the null and
cannot be adequately represented in simplified planar geometries.

We are also concerned with reconnection that occurs in the absence of
a null. The key entity in this case is the geometry of the
quasi-separatrix layers (QSLs), which provide a region of rapid
variation in the field line connectivity \citep{Demoulin-etal-1996,
  Priest-Titov-1996, Galsgaard-2000, Aulanier-etal-2006,
  Baker-etal-2009}. This behaviour can be quantified in terms of the
squashing factors of the configuration \citep{Titov-2007}, which give
a measure of the connectivity gradient.  Notably, MHD simulations of
line-tied QSL configurations suggest that, by systematically
decreasing the resistivity, current accumulation and the collapse to
small length scales may be unbounded \citep{Aulanier-etal-2005,
  Effenberger-etal-2011}. 

A related problem was tackled by \citet[][hereafter
Paper~I]{Craig-Effenberger-2014}. There, a non-resistive, Lagrangian
magneto-frictional relaxation method \citep{Craig-Sneyd-1986,
  Pontin-etal-2009, Craig-Pontin-2014} was used to obtain a near
singular relaxed configuration. It was found that the divergent
scaling of the QSL current could be significantly accelerated by the
presence of a magnetic null within the computational domain. QSL
diagnostics have also been used in related contexts, for example in
studies on tearing modes with particle in cell methods
\citep{Finn-etal-2014} or reduced MHD models of flux tube reconnection
in solar coronal loops \citep{Milano-etal-1999}.

The goal of the present study is to extend Paper~I by performing {\it
  resistive} three-dimensional MHD simulations with the finite volume
code PLUTO \citep{Mignone-etal-2007} to quantify the resistive
scalings of the simulated current layer. In common with Paper~I, we
show how the current build up is altered by shifting the position of a
magnetic null in the computational domain. To support our results, we
first summarize in Section~\ref{sec:scalings} theoretical arguments for
different resistive scalings. We then describe our field setup in
Section~\ref{sec:setup} before presenting our results and conclusion in the
last two sections. Details on the numerical implementation are given
in the Appendix.

\section{Theoretical Arguments for Resistive Scalings}
\label{sec:scalings}
Magnetic reconnection studies often assume a steady state description
based on an ``open'' geometry in which plasma, continuously washed
into the current layer, is ejected by the reconnection exhaust. This
approach generally leads to slow reconnection rates---Sweet-Parker
scalings possibly enhanced by flux pile-up factors---no matter whether
2D or 3D geometries are considered \citep{Craig-Fabling-1996}.  An
alternative method is to examine the dynamic collapse of ``closed''
line-tied X-points, which are subject to an initial,
topology-altering disturbance. In these models cold plasma is
contained within a highly conducting, rigid boundary across which
there is no mass flow. Under these conditions the reconnection rate $
\eta J $ can be fast---effectively independent of the plasma
resistivity $ \eta $. This scaling contrasts markedly with the
relatively weak build-up of current $J$ in Sweet-Parker merging $ \eta
J \sim \eta ^ {1 / 2} $.

Although the study of 3D X-point geometries is an active research
area, we know of no analytic argument that predicts the reconnection
rate associated with collapsing, 3D compressible X-points. A
linearized treatment is provided by \citet{Rickard-Titov-1996}, but
most studies rely on extrapolating numerical results, obtained for
computationally accessible resistivities (typically
$10^{-4}<\eta<10^{-2}$), down to physically plausible values
$\eta\lsim 10^{-9}$. Note that in the non-dimensional units we employ
(see below) the Alfv\'en crossing time of the system is of of order
unity and $\eta$ is an inverse Lundquist number.

One possibility for gaining insight into 3D reconnection scalings is
to examine the non-linear behaviour of planar X-point models under
the idealisation $ \partial / \partial z = 0$ but in the presence of
an axial ``guide'' field $ b \, \hat {\bf z} $. The common ground
between 2D and 3D X-point collapse models
\citep[\emph{e.g.}][]{Pontin-Craig-2005} also suggests that it is worthwhile
revisiting the non-linear $2\frac{1}{2}$D collapse problem.

\subsection {Planar X-point collapse, Y-point scaling}
We consider a planar potential field, immersed in a uniform background
plasma. As in Paper~I, we assume that length scales, densities and
magnetic field intensities are scaled with respect to typical coronal
values. Since we assume the gas pressure term to be negligible,
all wave motions are purely Alfv\'enic. All velocities are expressed
in units of the reference Alfv\'en speed $v_\mathrm{A}$. Time is measured in
units of the global length scale divided by $v_\mathrm{A}$.

The initial field can then be written
\begin{equation} 
{\bf B}_\mathrm{E} = \nabla \psi_\mathrm{E}  \times  \hat {\bf  z}   + b \, \hat {\bf  z}
\label{eq:xpoint}
\end{equation}
where $b \hat {\bf  z}$ is constant and
\begin{equation} 
\psi_\mathrm{E}  \, =   \frac{1}{2}(y^2-x^2) = - \frac{r^2}{2} \cos(2 \phi) 
\end{equation}
is the equilibrium flux function. For the moment we assume the
axial field is turned off, so $b = 0$.

The equilibrium field is line-tied on the outer boundary $r = 1$ and
subject to an initial radial wave disturbance $ \psi = \psi (r) $.
This launches an azimuthal disturbance field $ B_\phi = {\partial \psi
}/ {\partial r}$ in the form of a cylindrical wave that propagates
inwards from the outer boundary $ r = 1 $.  The imploding wave
continually steepens due to the gradient in the Alfv\'en speed $ v_\mathrm{A}
\propto r $.  However, a point $ r = R $ (say) is reached where the
disturbance field begins to overwhelm the background field.
Cylindrical symmetry is then lost as the X-point field is weakened
or reinforced in adjacent lobes. The disturbance field now becomes
increasingly one-dimensional $ B_\phi \to B_\mathrm{s} $ and, in the absence of
resistivity, a Y-type, finite time singularity emerges.  For finite
$ \eta $, however, a resistive length scale $ \Delta $ is introduced
that allows the wave speed $ B_\mathrm{s} / \sqrt{\rho} $ to be matched by the
diffusion speed $ \eta/ \Delta $.  In this case, by neglecting the
relatively weak density dependence of the wave speed, we obtain
\citep{McClymont-Craig-1996}
\begin{equation}
\Delta  \sim \eta,   \quad  R \sim B_\mathrm{s} \sim \eta ^0,  \quad J \sim \eta ^{-1}.
\label{eq:theo-scaling}
\end{equation}
These resistive scalings define the thickness, length, field strength
and current density of the Y-point field.  Note that the length
$R\gg\Delta$ of the current sheet is determined by the radius at which
cylindrical symmetry is lost and generally lies well outside the
diffusion layer of thickness $\eta$.

\begin{figure}[h]
  \centering 
 \includegraphics[width=0.49\textwidth]{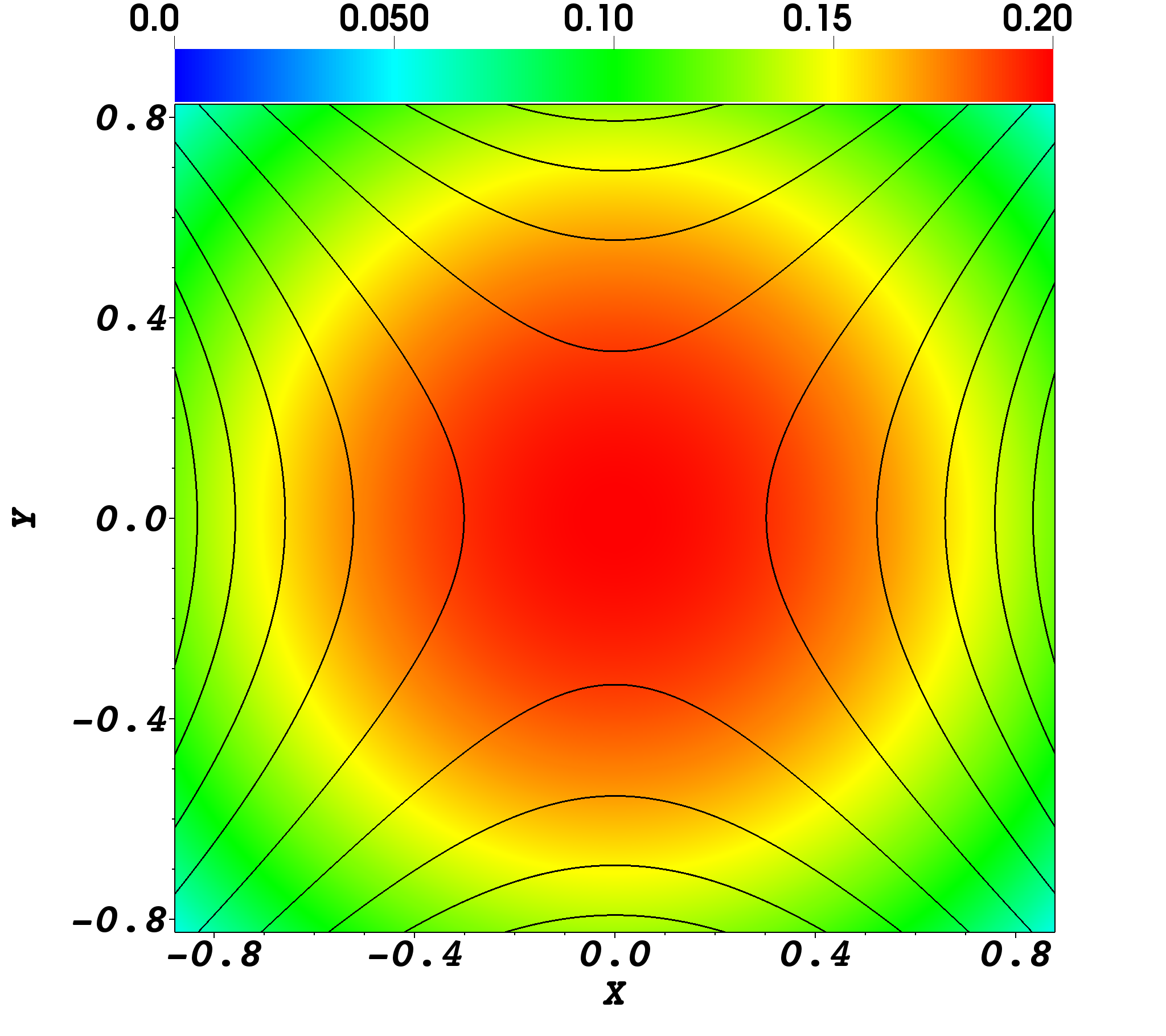}
 \includegraphics[width=0.49\textwidth]{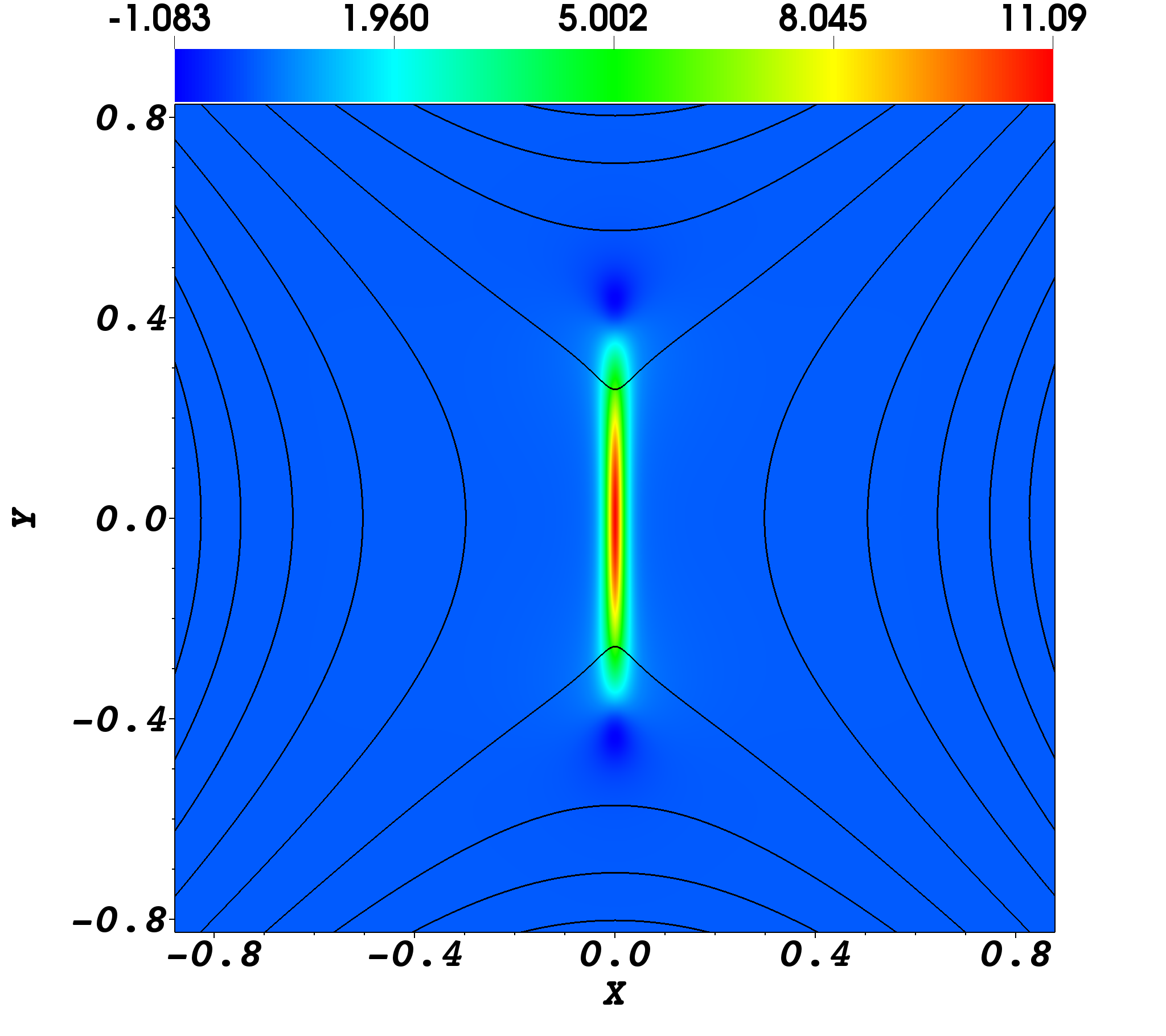}
  \caption{Illustration of the two-dimensional Y-point current
    formation and field line structure. The color coding gives the
    $J_z$ current while the black lines show field lines, integrated
    from the boundary. The left panel shows the initial condition at
    $t=0$, while the right panel is a snapshot at $t=2$ (Alfv\'en
    times). The run was performed on a grid of $400\times 400$ points
    with $\eta=3\times 10^{-3}$.}
  \label{fig:ypoint}
\end{figure}
To illustrate the topology and current formation process of the
previous discussion, we performed a simple two-dimensional, line-tied
simulation of the X-point field of Equation~(\ref{eq:xpoint}) with
perturbation $\psi_\mathrm{p} = A(1-x^2)(1-y^2)$. In this case we find that
$A=0.05$ is sufficient to provide a current layer of global length
consistent with the scalings of Equation~(\ref{eq:theo-scaling}) and
the non-linearity condition $ R \gg \Delta $.ñ

Here, we only present two snapshots of the evolution in
Figure~\ref{fig:ypoint}. While initially, the configuration is nearly
symmetric, at a later stage, due to the perturbation and the
corresponding change in topology, the current collapses to a nearly
one-dimensional structure and the magnetic field lines adapt to the
Y-point geometry around the endpoints of the current
sheet. Reduction of the resistivity leads to correspondingly larger
current densities and smaller perpendicular length scales of the
sheet. The perturbation amplitude $A$ determines mainly the length of
the current sheet, but, within limits set by the global geometry, does
not influence resistive scalings for the peak current density and
current sheet thickness.  We do not pursue the investigation of this
2D configuration further at this point, since this has already been
done extensively in previous studies \citep[\emph{e.g.},][and
references therein]{Priest-Forbes-2000}. More details on the linear
and non-linear treatment of X-point plasmas can be found for example
in \citet{Craig-McClymont-1993}.

The scalings of Equation~(\ref{eq:theo-scaling}) are based on an
analysis given by \citet{McClymont-Craig-1996}. These authors point
out that these results are in good agreement with dynamic simulations
of resistive current layers as well as the formally exact, non-linear
imploding field models of \citet{Forbes-1982}, which yield $\Delta
\sim \eta^{0.892}$ and $J\sim\eta ^{-1.04}$. Even so, the fast reconnection
rate is expected to stall when significant back pressures due to
strong axial fields oppose the localization.

\subsection {Arguments for QSL Scalings}
Suppose now that the axial field is no longer vanishing. The null
point is removed but the axial field is compressed by the implosion
leading to back pressures which oppose the localization.
\citet{McClymont-Craig-1996} show that the fast scaling can persist
only if
\begin{equation}
\frac {1} {2} b^2 \le \eta.
\label{eq:fastscale}   
\end{equation}
When this condition is not met the reconnection rate can become very
slow, approaching the static limit $ \eta J \sim \eta $.

We are primarily interested in how these results impact on 3D nulls in
the presence of QSLs. Clearly, insofar as QSL structures can be
represented by X-points threaded by axial fields, they cannot be
expected to provide fast reconnection, at least for realistic
amplitudes $ b $.  In the case of fully 3D line-tied configurations,
however, the presence of a null at some point within the source volume
should act to focus and intensify the current. This tendency could
be reinforced by the strong squashing factors that characterize the
QSL configuration. Our expectation, therefore, is that while
the reconnection rate associated with strong QSLs is probably slow, it can
be significantly accelerated if a null point is located in the
vicinity of the QSL. This reasoning is explored computationally in
the analysis that follows.
   
\section{Potential Field Formulation}
\label{sec:setup}
We base our analysis on the potential field model of Paper~I [Equation~(9)], 
which has the general form
\begin{eqnarray}
\mathbf{P} = [x, (\mu-1)y, -\mu z +b] \,.
\label{eq:pot-field}
\end{eqnarray}

The field is defined within the cubic domain $x,y,z=\pm 1$, and
line-tied on the bounding surfaces. Note that although $\mu=0$ allows
X-point fields typified by Equation~(\ref{eq:xpoint}) to be modelled,
we now have line-tied boundary conditions at $z=\pm 1$. More
generally, $\mu$ can be regarded as a proxy for the field asymmetry,
which we fix at $-0.4$ in the present study. The parameter $b$ is more
significant since it allows a continuously shift of the magnetic null
along the $z-$axis. The null is located at the point
$\mathbf{r}_0=(0,0,b/\mu)$ and, by adjusting $b$, the null can be
transferred outside the computational domain. Thus, we find nulls
located at $z_0=-0.75$ and $z_0=-1.25$, corresponding to the
respective choices $b=0.3$ and $b=0.5$. The first case provides a
convenient platform for null point reconnection to be modelled.
However, by taking $b=0.5 $ the null ``disappears'' while the QSL
geometry is retained, potentially altering the reconnection rate. We
note that the type of potential field considered here can be regarded
as an approximation to a more general force free field; see the
discussion in Paper~I.

To initiate the current formation, a suitable perturbation of the
equilibrium field $\mathbf{P}$ is required. The combined initial field
configuration is then given by $\mathbf{B}_0 = \mathbf{P} +
\mathbf{B}_\mathrm{p}$ and we assume the initial velocity field to vanish. For
comparison with Paper~I, we adopt the same type of perturbation field,
with amplitude $A=0.3$:
\begin{eqnarray}
\mathbf{B}_\mathrm{p} = [A \sin(\pi x/2)\cdot(1-y^2)\cdot(1-z^2)\cdot \exp(-4 x^2 -3 y^2)]\hat\mathbf{y} \,.
\end{eqnarray}
In the vicinity of the origin this perturbation reduces to the form
${\bf B}_\mathrm{p} \propto (0, x, 0)$. As shown in Paper~I, when
added to the the X-point field of Equation~(\ref{eq:pot-field}), the
effect is to tilt the spine and drive implosive currents within the
fan. The global perturbation also contains additional exponential
factors whose role is to constrain initial currents and forces in the
outer field to be of order unity. Their slightly different strengths
are chosen to break any artificial symmetries imposed by the
perturbation. Further discussion on the form of the perturbation can
also be found in Section 2.6 of \citet{Craig-Pontin-2014}. Details of
the numerical setup which is used to solve the field evolution are
given in the Appendix.

It should be stressed that the role of the idealised global
perturbation is solely to initiate a collapse towards a resistive
current layer. For actual coronal fields, the perturbation is probably
supplied by photospheric motions during a slow build-up phase, while
we impose the perturbation directly with the initial field in the
domain. Other studies \citep{Galsgaard-2000, Aulanier-etal-2005,
  Effenberger-etal-2011} include such motions by suitable velocity
fields at the boundaries. To retain the direct comparability with our
previous results from Paper~I and to keep the run times for individual
simulations relatively short, we stick with this kind of direct
perturbation field. The small divergences of the magnetic field
introduced by this particular form of perturbation ---or more
generally through discretisation errors in the code--- were not
relevant in the study of Paper~I since the Lagrangian scheme is
solenoidal by construction. In the present study, the smallness of the
perturbation and the divergence cleaning method keep the divergence
error low. The scaling results discussed in the next section should
not depend on the actual form chosen for the perturbation and its
amplitude (\emph{cf.} the previous discussion on the planar X-point
collapse).

\section{Current Structure, Time Evolution, and Scaling Results}
\label{sec:results}

We performed runs of the two field configurations described in the
previous section ($b=0.3$ and $b=0.5$) for different values of $\eta$,
between $10^{-2}$ and $10^{-4}$. Since previous analytic and numerical
studies have confirmed that the properties of the current layer at
peak current \citep{Heerikhuisen-Craig-2004} provide a reliable guide
to the reconnection rate, we follow the magnetic field evolution until
maximum current density is achieved.

For larger resistivities, the problem is too diffusive to allow for a
sufficiently strong current localization, so we disregard results from
such runs. Similarly, resistivities smaller than $10^{-4}$ turn
out to be computationally unfeasible, since even at the highest
transverse resolution in the current sheet region of $\Delta = 5\times
10^{-4}$ (\emph{cf.} the description of the numerical grid in the appendix)
the numerical dissipation will start to dominate the evolution.

\begin{figure}[h]
  \centering 
 \includegraphics[width=0.49\textwidth]{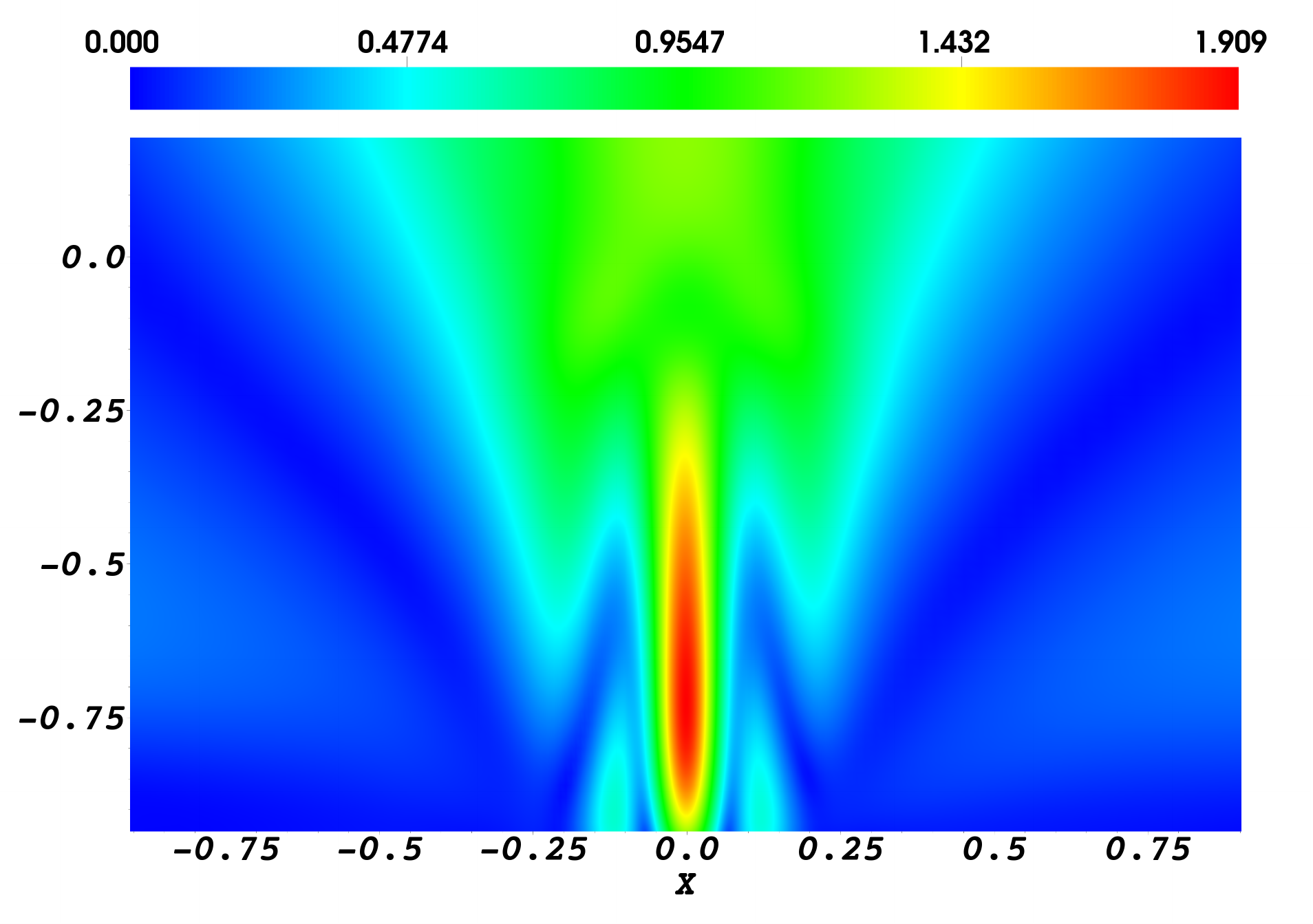}
 \includegraphics[width=0.49\textwidth]{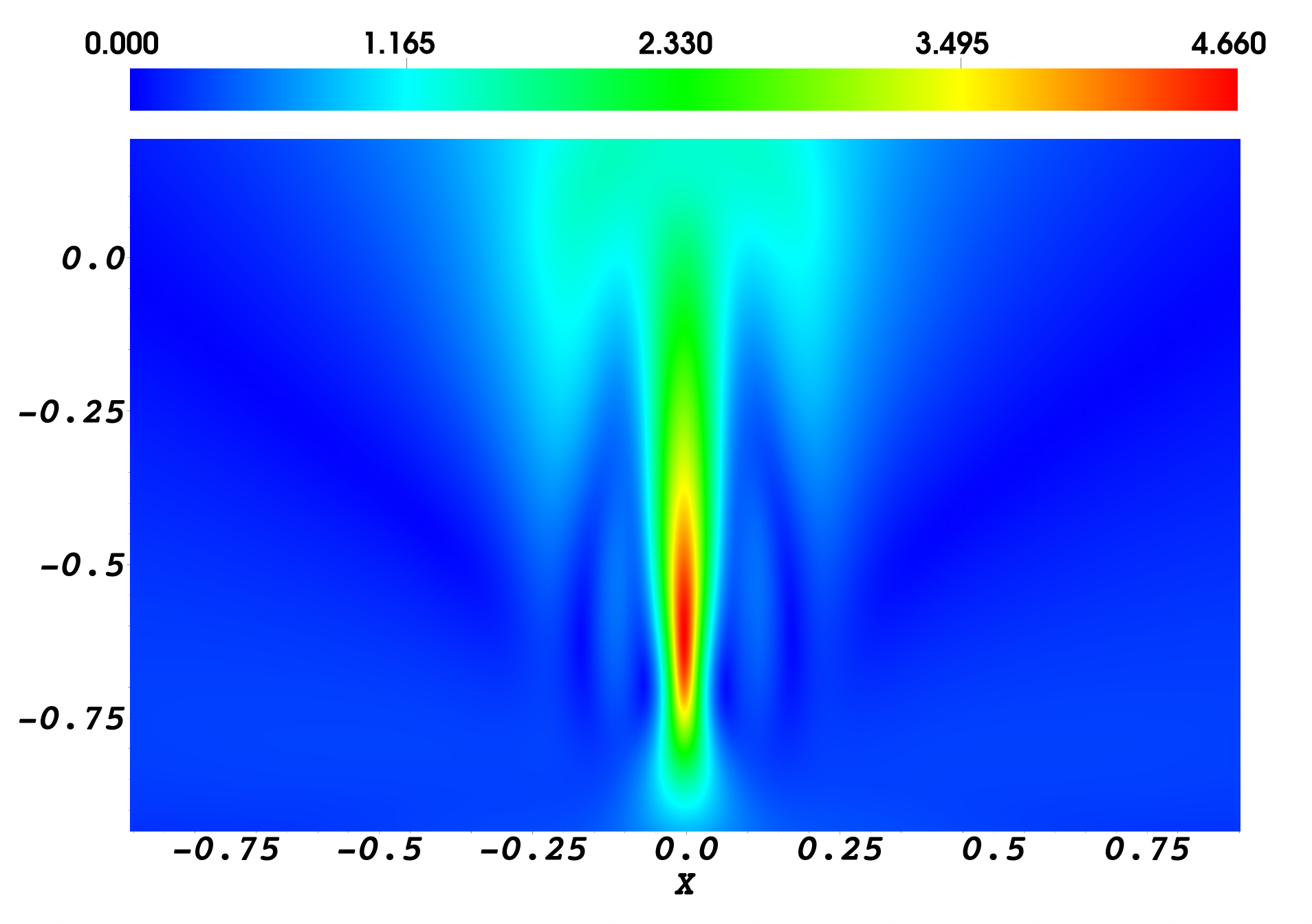}\\
 \includegraphics[width=0.49\textwidth]{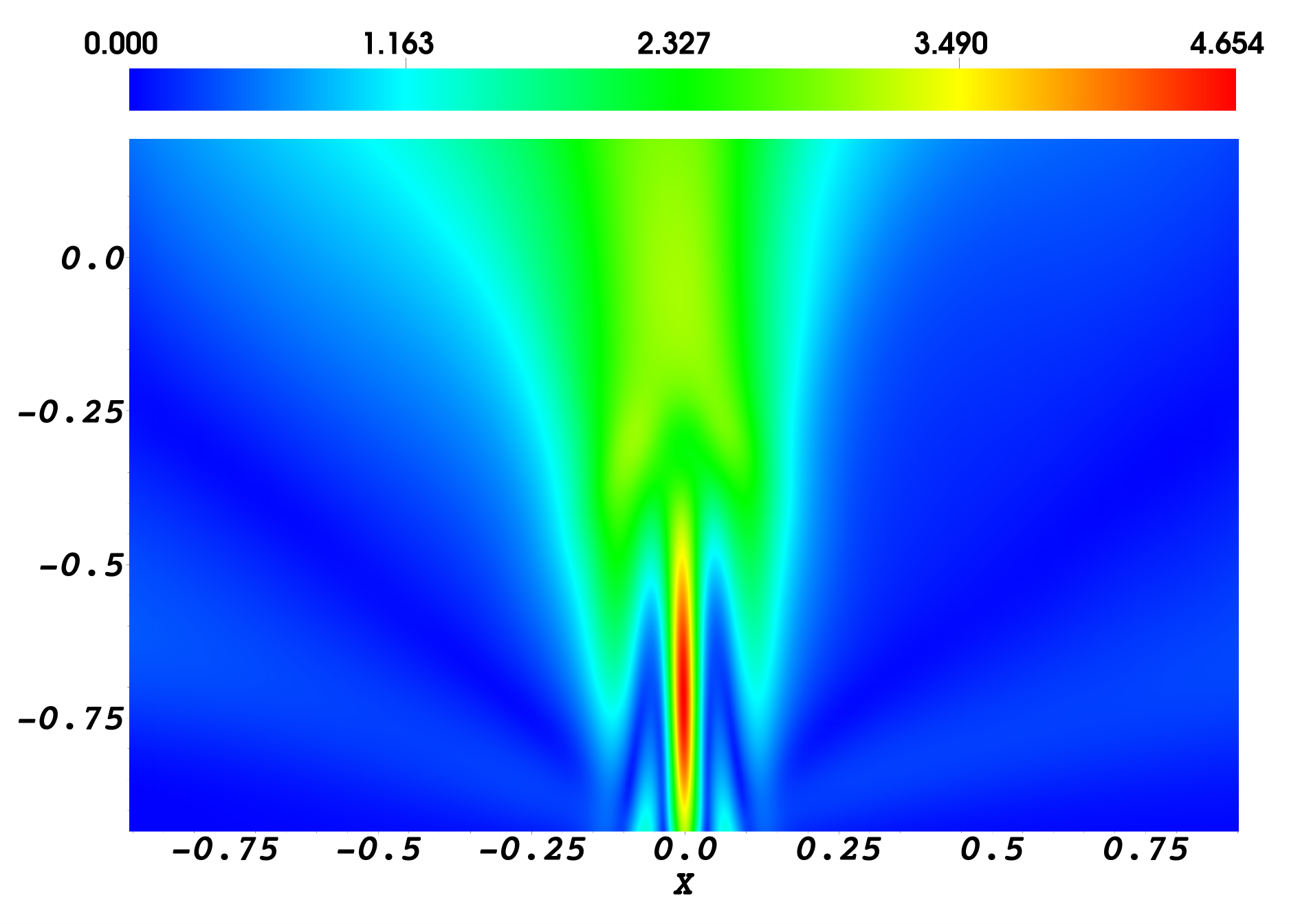}
 \includegraphics[width=0.49\textwidth]{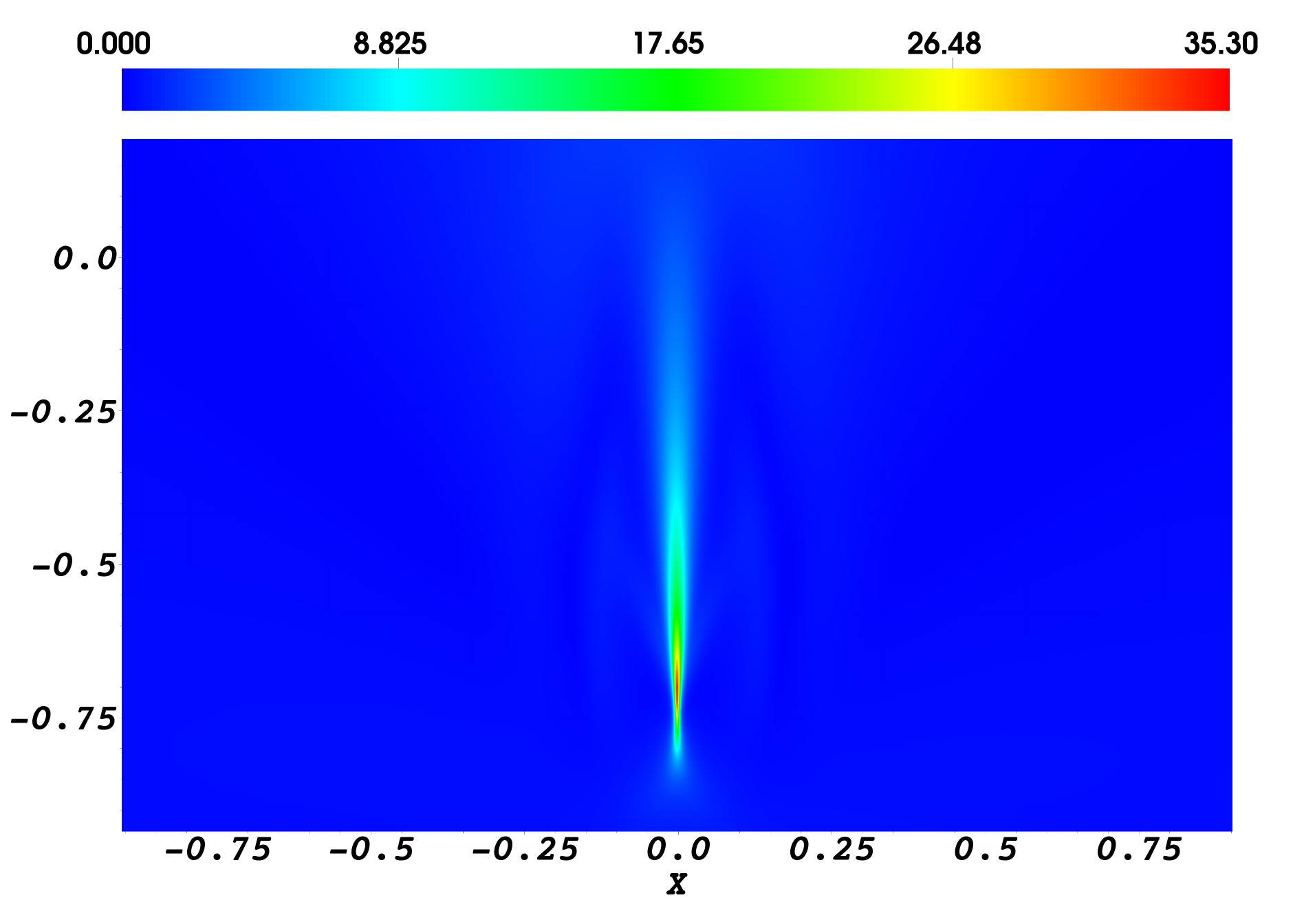}
  \caption{Current sheet structure around the time of peak current
    density in the $x$-$z$-plane ($y=0$); the color coding gives the current
    density. The left column shows results for the QSL only
    configuration ($b=0.5$) while the right column shows the
    configuration with a null at $-0.75$ ($b=0.3$). The top row has a value
    for the resistivity of $\eta=10^{-3}$ and the bottom row
    $\eta=10^{-4}$.}
  \label{fig:currentstructure}
\end{figure}

Figure~\ref{fig:currentstructure} shows the structure of the
accumulated current in a planar cut at $y=0$ around the time of peak
current density. The left column gives results for the QSL-only
configuration ($b=0.5$) and the right column shows the current
structure when a null is present at $z=-0.75$ ($b=0.3$). The strong
localization provided by the null is clearly a dominant feature, while
the current remains more broadly distributed without the null. The
collapse to much smaller length scales for respectively smaller
resistivities is evident, and highlights the need for sufficient grid
resolution.

\begin{figure}[h]
  \centering 
  \includegraphics[width=0.99\textwidth]{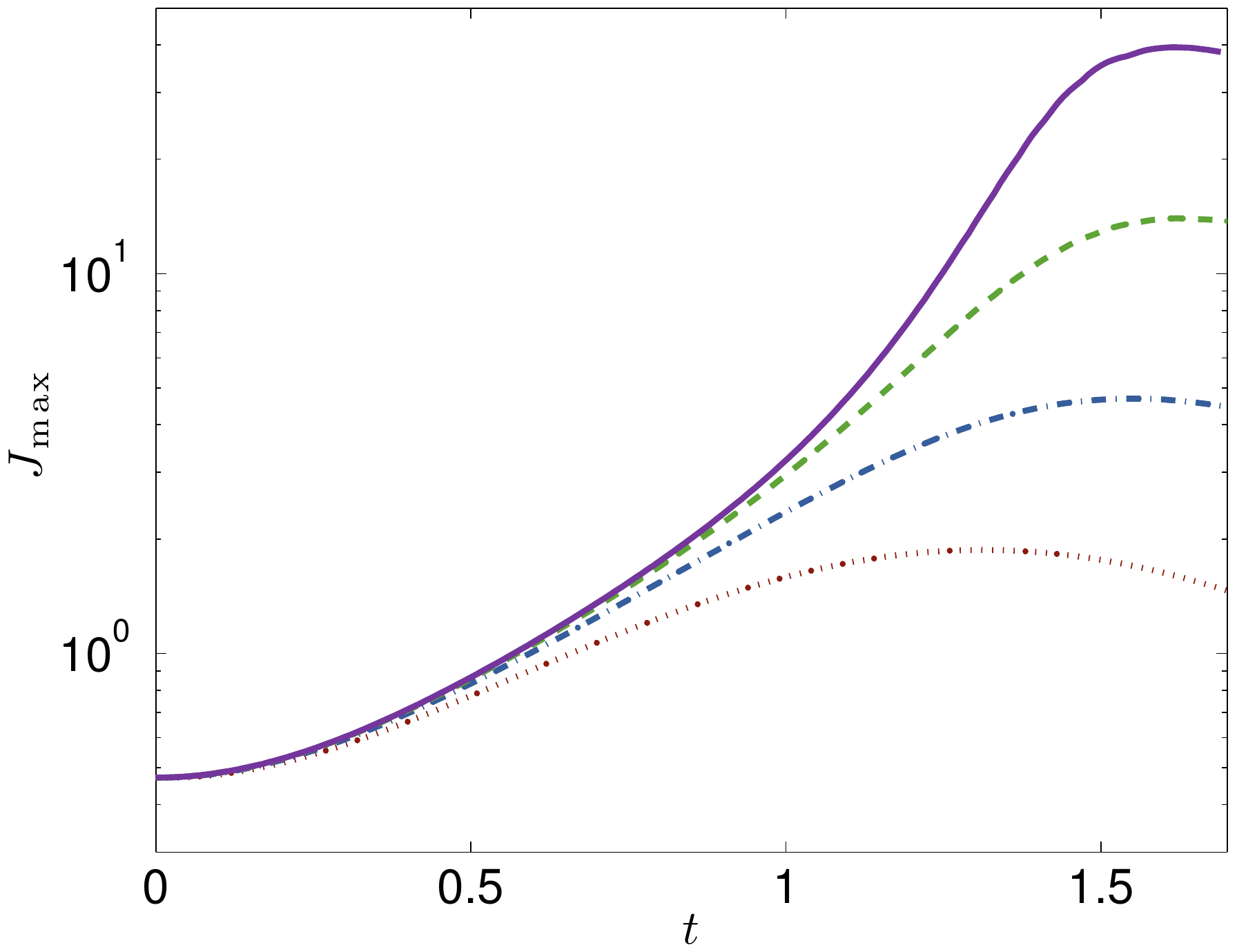}
  \caption{Time evolution of the maximum current build up in the
    central current sheet region in the magnetic null case ($b=0.3$) for
    four different resistivities, $\eta = 3\times 10^{-3}$ (red,
    dotted), $\eta = 10^{-3}$ (blue, dot-dashed), $\eta = 3\times
    10^{-4}$ (green, dashed), and $\eta = 10^{-4}$ (purple, solid).}
  \label{fig:timeevo}
\end{figure}
The time evolution of the current build-up in the null-point case is
given in Figure~\ref{fig:timeevo}, for four different
resistivities. As expected, the build-up takes longer for the more
intense localizations associated with the weaker resistivities.
Specifically, the maximum current density peaks at about 1.5 Alfv\'en
times for low resistivities, while for higher values, the evolution is
broader and less pronounced. It appears that a distinct fast phase of
current accumulation is only present for small resistivities, hinting
at a self-maintained sheet collapse, which is inhibited at too large
$\eta$. Presumably this is why resistive current layers at
sufficiently small resistivities very closely resemble the
near-singular, force-free, current structures that derive from the
magneto-frictional relaxation of Paper~I.

\begin{figure}[h] 
  \centering 
  \includegraphics[width=0.99\textwidth]{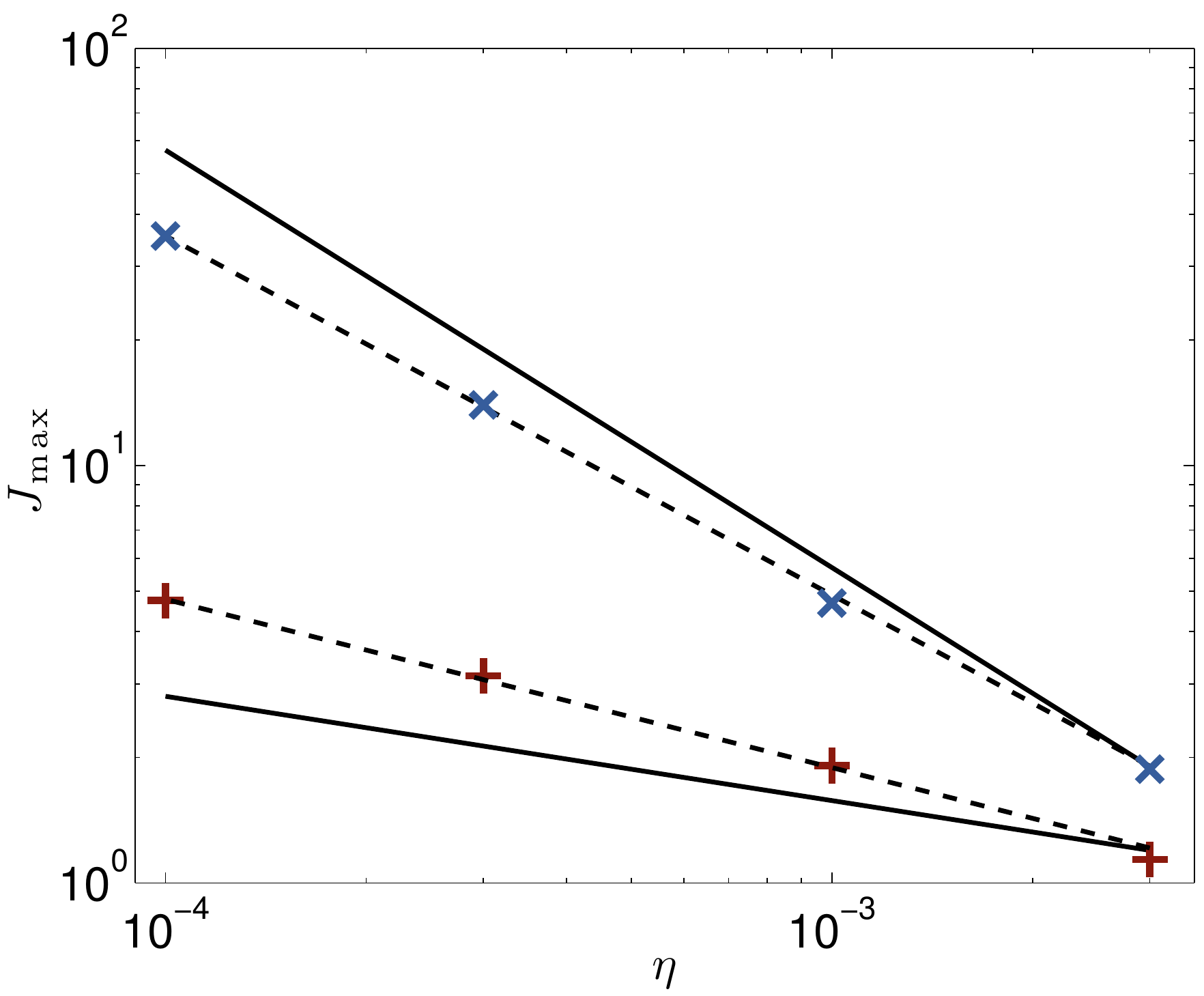}
  \caption{Resistive scalings of the peak current density for $b=0.3$
    (blue, x), \emph{i.e.}\ the magnetic null in the domain, and $b=0.5$
    (red, +), \emph{i.e.}\ only the QSL structure present. The dashed lines
    give the best-fit power laws ($J\sim\eta^{\gamma}$) with exponents
    $\gamma=-0.86$ and $\gamma=-0.40$, respectively. The solid lines
    represent theoretical limiting cases of a $\gamma=-1$ and $\gamma
    = -0.25$ scaling, to guide the eye.}
  \label{fig:scalings}
\end{figure}
Figure~\ref{fig:scalings} summarizes the main scaling results of our
study. Blue crosses indicate the peak current densities for runs with
different resistivities when the null is in the domain, while the red
$+$ symbols give the results for QSL only currents. We see a clearly
different trend in the scaling of maximum current with
resistivity. While the best-fit slope (dashed lines) is close to a
theoretical $\eta^{-1}$ scaling (see the discussion in
Section~\ref{sec:scalings}) when the null-point lies in the domain, the
scaling is much weaker, tending towards a weak scaling between
$\eta^{-0.5}$ and $\eta^{-0.25}$, when only the QSL structure is
present. Note that the lower bound of $\eta^{-0.25}$ follows from the
Lagrangian results of Paper~I, where a scaling exponent of about $0.5$
for the current accumulation with resolution was found. This converts
to an Eulerian exponent of $0.25$ in a strictly one-dimensional
setting \citep[see the discussion in the appendix of
][]{Craig-Litvinenko-2005}. In a three-dimensional configuration,
however, there is no exact equivalence between the Lagrangian and
Eulerian description, so this can only be regarded as a rough lower
bound. The results corroborate the earlier findings of Paper~I, in
that there is a qualitatively different current accumulation behavior,
induced by the three-dimensional magnetic null. We emphasize, however,
that we see no signs of saturation of current concentration in either
case, for the considered range of resistivities.

An interesting feature of the above results is the extent to which
they differ from a $2\frac{1}{2}$D description, based on a planar,
line-tied X-point threaded by a uniform axial field. As indicated by
Equation~(\ref{eq:fastscale}), and confirmed by numerical studies,
fast reconnection is easily stalled by the back pressure introduced by
the compression of $ b \,\hat {\bf z} $.  This problem is worsened
when line-tying on the upper and lower planes $ z = \pm 1 $ is
introduced, because tension of the axial field can come into play.  In
this case, modest axial fields can prevent the X-point
implosion \citep{Craig-Pontin-2014}.  Only by reducing the initial $ b \,
\hat {\bf z} $ amplitude, or extending the distance between the
$z$-boundaries, can the collapse be recovered. Clearly, the presence
of a null negates these difficulties.

Finally, we should mention that although gas pressure is negligible in
all our simulations, the extreme scalings required to maintain fast
reconnection are likely to remain sensitive to finite $ \beta $
effects \citep{Pontin-etal-2007}. Since coronal plasmas are low $ \beta
$, however, the principle features of X-point collapse should not be undone.

\section{Summary and Conclusion}
\label{sec:conclusion} 
We have studied the implosive collapse of a line-tied 3D X-point and
QSL configuration, using a resistive MHD code. Our results, evaluated
at the time of peak current density, directly compare resistive
scalings for perturbed QSL and null-point initial equilibrium
fields. The implication is that, while QSL equilibria lead to
divergent current structures in the limit of small $\eta$, the
reconnection rate is considerably enhanced by the presence of a null
within the computational domain. Notably, for moderate finite
amplitude disturbances, the merging rate can approach the fast scaling
$ J \sim \eta ^ {-1} $ over the range of resistivities considered. In
contrast, the QSL scalings, though still divergent, are significantly
weaker, \emph{i.e.}\ about $ J \sim \eta ^ {-0.4}$.

It is interesting that the present results provide a direct extension
of the results provided by the Lagrangian magneto-frictional method of
Paper~I. In the Lagrangian approach, $\nabla \cdot {\bf B}$ is
guaranteed to vanish and flux conservation is automatically
satisfied. The proper time evolution is not accessible, however, and
results have to be extracted from the near-singular, force free-field
that emerges during the late stages of the evolution. That the methods
of Paper~I and the resistive MHD approach provide a consistent
physical interpretation goes some way, we believe, in establishing the
veracity of the present results. It will be of interest therefore, to
explore these findings using additional parallel computing capacities
to improve the numerical resolution and extend the parameter range of
the simulations. An extension to different cases of boundary
conditions and perturbations by boundary driving should yield new
insights on the general applicability of our findings to different
solar flare scenarios.

\begin{acks}
  We acknowledge the work that has been devoted to the development and
  documentation of the PLUTO MHD code used in this study. Work
  performed by F.E. was partially supported by NASA grant
  NNX14AG03G. Constructive comments of an anonymous referee are
  appreciated.
\end{acks}

\newpage
\appendix   
\section{Code Setup}
We evolve the perturbed field of Section~\ref{sec:setup} according to
the full set of non-relativistic MHD equations, using the finite
volume code PLUTO (version 4.1, see
http://plutocode.ph.unito.it). The code manual and the
foundational papers of the code \citep{Mignone-etal-2007,
  Mignone-etal-2012} describe tests and the fundamental capabilities
and solution methods available in the code. Here, we only briefly
describe the setup choices and configuration we use for our particular
problem.

The ideal MHD equations are implemented in the code in a conservative
form, given by
\begin{eqnarray}
  \partial_t \rho &+& \nabla \cdot (\rho {\bf v})   = 0   \label{eq:conti}\\
  \partial_t (\rho {\bf v}) &+& \nabla \cdot \left[\rho {\bf v v}
          - {\bf B B} +  p_\mathrm{t} {\bf I}    \right] = 0   \label{eq:mom}\\
  \partial_t {\bf B} &+& \nabla \cdot ({\bf vB-Bv}) = 0   \label{eq:indu}
\end{eqnarray}
where $\rho$ is the mass density, ${\bf v}$ is the velocity and ${\bf
  B}$ the magnetic field. The total pressure is written as $p_\mathrm{t} = p +
{\bf B}^2/2$ and ${\bf I}$ is a unit tensor.

We use the simple isothermal equation of state, where the pressure is
given by $p=\rho c_s^2$, and the sound speed is set to a small value
of $c_\mathrm{s}=10^{-3}v_\mathrm{A}$ to only have negligible compressive effects in the
evolution. Viscous and resistive dissipation effects are explicitly
added as parabolic terms on the right-hand side of the conservation
equations to control the non-idealness of the evolution in relation to
numeric dissipation. We always choose the viscosity $\nu$ to be equal
to the resistivity $\eta$ and uniform across the domain. The
time-evolution is implemented as second-order Runge-Kutta
time-stepping and we use the Roe type Riemann solver to minimize
numeric dissipation. The divergence free constraint of the magnetic
field is maintained by the divergence cleaning method
\citep{Dedner-etal-2002}. We checked for various times during the
evolution that the method is successful in keeping the divergence
error equal or below 1 percent of the current magnitude.

To keep numeric diffusivity small against the explicit dissipation
terms, a sufficiently high grid resolution is needed in the vicinity
of the strong current layers. To achieve this feat with limited
computational resources, we make use of the stretched grid features of
PLUTO. We decompose the domain in the transverse $x$ and $y$
directions into three subdomains. From $-1$ to $-0.1$ we have a
stretched grid of decreasing cell size, matching to the uniform grid
from $-0.1$ to $0.1$ in the sheet region. From $0.1$ to $1$ the cell
size increases again. Each of the outer subdomains is covered by 50
grid points, while we have up to 400 points in the centre region,
giving a maximum resolution (or grid-cell size) of $d=5\times 10^{-4}$,
which we use for runs with the smallest resistivities. The resolution
in $z$ direction is kept constant at 100 points, assuming that the
relevant dynamic scales are only perpendicular to the $z$-axis. We
confirmed the soundness of our grid approximation with comparison runs
between uniform grids and the described stretched patch grid. We found
the differences in field and current magnitudes to be smaller than a
few percent, and the qualitative field evolution was practically
indistinguishable, encouraging our confidence in the choice for the
grid setup.




\tracingmacros=2
\bibliography{effenberger-craig-2015-solphys}  

\IfFileExists{\jobname.bbl}{} {\typeout{}
\typeout{****************************************************}
\typeout{****************************************************}
\typeout{** Please run "bibtex \jobname" to obtain} \typeout{**
the bibliography and then re-run LaTeX} \typeout{** twice to fix
the references !}
\typeout{****************************************************}
\typeout{****************************************************}
\typeout{}}

\end{article} 

\end{document}